\documentclass[reprint,amsmath,amssymb,aps,prb,showpacs,floatfix,byrevtex]{revtex4-1}

\usepackage{amsfonts}
\usepackage{bm}
\usepackage{dcolumn}
\usepackage{graphicx}
\usepackage{wasysym}

\begin{document}

\preprint{Draft --- not for distribution}

%
%
%
\title{Optical properties of the iron-arsenic superconductor \boldmath BaFe$_{1.85}$Co$_{0.15}$As$_2$ \\
  \unboldmath }
\author{J. J. Tu}
\author{J. Li}
\author{W. Liu}
\author{A. Punnoose}
\affiliation{Department of Physics, The City College of New York, New York, NY 10031}
\author{Y. Gong}
\author{Y. H. Ren}
\affiliation{Department of Physics and Astronomy,
 Hunter College of the City University of New York, New York, NY 10065}
\author{L. J. Li}
\author{G. H. Cao}
\author{Z. A. Xu}
\affiliation{Department of Physics, Zhejiang University, Hangzhou 310027, China}
\author{C. C. Homes}
\email{homes@bnl.gov}
\affiliation{Department of Condensed Matter Physics and Materials Science, Brookhaven National Laboratory, Upton, NY 11973}

\date{\today}

%
%
\begin{abstract}
The transport and complex optical properties of the electron-doped iron-arsenic
superconductor BaFe$_{1.85}$Co$_{0.15}$As$_2$ with $T_c = 25$~K have been examined
in the Fe-As planes above and below $T_c$.   A Bloch-Gr\"{u}neisen analysis of
the resistivity yields a weak electron-phonon coupling constant $\lambda_{ph}\simeq 0.2$.
The low-frequency optical response in the normal state appears to be dominated
by the electron pocket and may be described by a weakly-interacting Fermi liquid
with a Drude plasma frequency of $\omega_{p,D} \simeq 7840$~cm$^{-1}$
($\simeq 0.972$~eV) and scattering rate $1/\tau_D \simeq 126$~cm$^{-1}$
($\simeq 15$~meV) just above $T_c$. The frequency-dependent scattering rate
$1/\tau(\omega)$ has kinks at $\simeq 12$ and 55~meV that appear to be related to
bosonic excitations.
Below $T_c$ the majority of the superconducting plasma frequency originates from the
electron pocket and is estimated to be $\omega_{p,S}\simeq 5200$~cm$^{-1}$ ($\lambda_0
\simeq 3000$~\AA ) for $T\ll T_c$, indicating that less than half the free carriers
in the normal state have collapsed into the condensate, suggesting that this material
is not in the clean limit.  Supporting this finding is the observation that this material
falls close to the universal scaling line for a BCS dirty-limit superconductor
in the weak-coupling limit.
There are two energy scales for the superconductivity in the optical
conductivity and photo-induced reflectivity at $\Delta_1(0) \simeq 3.1\pm
0.2$~meV and $\Delta_2(0) \simeq 7.4\pm 0.3$~meV.  This corresponds to
either the gapping of the electron and hole pockets, respectively, or
an anisotropic {\em s}-wave gap on the electron pocket; both views are
consistent with the $s^\pm$ model.

\end{abstract}
%
%
%
%
%
%
%
%
\pacs{74.25.Gz, 74.70.Xa, 78.30.-j}%
\maketitle

%
%
\section{Introduction}
Since its discovery nearly a century ago, the field of superconductivity has
periodically reinvented itself.  A significant advance in the understanding of
this phenomenon was achieved through the model by Bardeen, Cooper and Schrieffer
(BCS), which proposed that below a critical temperature $T_c$ electrons condense
into pairs which are coupled through lattice vibrations;\cite{bcs} these so-called
Cooper pairs are bosons which constitute a supercurrent that may flow without
loss.  Below $T_c$ an isotropic {\em s}-wave gap also opens in the spectrum of
excitations across the Fermi surface.  Within this model, it was thought that the
$T_c$'s of conventional metals and alloys could not exceed $\simeq 30$~K (Ref.
\onlinecite{mcmillan68}).  The discovery of superconductivity at elevated
temperatures in the copper-oxide materials\cite{bednorz86} with $T_c$'s in
excess of 130~K at ambient pressure and an unusual {\em d}-wave superconducting
energy gap\cite{harlingen95} indicates that these materials are not phonon mediated;
indeed, the coupling mechanism in these materials remains unresolved and a subject
of considerable debate.  The discovery  of superconductivity in MgB$_2$ with a
surprisingly high transition temperature $T_c = 39$~K intially suggested an
unusual pairing mechanism.\cite{nagamatsu01}  However, the isotope effect
firmly established that this material is phonon mediated;\cite{budko01} in
this case the high phonon frequencies in MgB$_2$ are likely responsible for
the enhanced critical temperature.\cite{kortus01}
%
%

%
%
The recent discovery superconductivity in the iron-arsenic LaFeAsO$_{1-x}$F$_x$
(1111) compound\cite{kamihara08} at ambient pressure ($T_c = 26$~K)
was surprising because iron and magnetic impurities in general were considered
detrimental to the formation of superconductivity; $T_c$'s $\gtrsim 50$~K
in this material were quickly achieved through rare-earth
substitutions.\cite{renz08a,renz08b,wang08,ishida09}
While such high values for $T_c$ do not necessarily rule out a phonon-mediated
pairing mechanism, the close proximity of magnetic order and superconductivity in
these compounds\cite{cruz08} has lead to the suggestion that the pairing in this
class of materials may have another origin.\cite{boeri08,boeri09,mazin10,johnston10}
While the highest $T_c$'s are observed in the 1111 compounds, large single
crystals of these materials have proven difficult to grow.  For this reason
attention has shifted to the structurally-simpler BaFe$_2$As$_2$ (122) materials,
where large single crystals are readily available.  Unlike the cuprates, the
parent compound is metallic with a structural and magnetic transition at $T_N
\simeq 140$~K, below which it remains metallic.\cite{rotter08a}  The emergence
of superconductivity and the suppression of the magnetic and structural
transitions may be achieved through the application of pressure\cite{alireza09}
or through chemical substitution;\cite{renz09,rotter08b,sefat08,li09} K-doping
results in a hole-doped material with a maximum $T_c \simeq 38$~K, while
Ni- and Co-doping result in electron-doped materials with a somewhat
lower maximum $T_c \simeq 29$~K.
%
%
The electron- and hole-doped materials have been the subject of numerous
investigations, including thermodynamic and transport studies,\cite{rotter08b,sefat08,ren08a,%
li09,fang09,park09,yuan09,martin09,budko09,rullier09,yashima09,gofryk10,ning10,tanatar10,reid10}
angle-resolved photoemission\cite{ding08,liu08,khasanov09,evtushinsky09,koitzsch09,zabolotnyy09,%
terashima09,sekiba09,liu10,nakayama10} (ARPES),
Raman\cite{litvinchuk08,chauviere09,rahlenbeck09,muschler09} and optical
studies.\cite{li08,hu09,yang09,wu09a,fischer10,heumen09,kim10,gorshunov10,%
nakajima10,wu10a,marsik10,perucchi10,barisic10,dressel10,lucarelli10,lobo10}
Several of these studies address the nature of the superconducting gaps,
such as whether there are multiple gaps and
whether the gaps contain nodes.  The experimental picture is not clear.  The
isotope effect in these materials is also uncertain; while a large Fe isotope
effect has been reported in BaFe$_2$As$_2$  and Ba$_{0.6}$K$_{0.4}$Fe$_2$As$_2$
polycrystals,\cite{liu09} an inverse iron isotope effect has also been
observed.\cite{shirage09}

%
%
Despite the structural differences of the iron-arsenic compounds, the
band structure of these materials is remarkably similar, with a minimal
description consisting of a hole band ($\alpha$) centered at the $\Gamma$
point and an electron band ($\beta$) at the $M$ point of the
Brillouin zone.\cite{raghu08,singh08a,singh08b}  The mobilities are thought to be
significantly higher in the electron bands.\cite{rullier09}
Several theoretical models\cite{mazin08,kuroki08,maier08,chubukov08} propose
that below $T_c$ superconducting {\em s}-wave energy gaps form on the electron
and hole pockets, possibly with a sign change between them, the so-called
$s^\pm$ model.  In the $s^\pm$ model the gap on the electron pocket may be an
extended {\em s}-wave with nodes on its Fermi surface;\cite{chubukov09}
it is possible that disorder may lift the nodes.\cite{mishra09}

%
%
In this work the electron-doped BaFe$_{1.85}$Co$_{0.15}$As$_2$ superconductor
($T_c = 25$~K) has been examined using several different techniques.  The
temperature dependence of the {\em ab}-plane resistivity is analyzed using
a simple Bloch-Gr\"{u}neisen framework, yielding an upper-bound for the
electron-phonon coupling constant of $\lambda_{ph} \simeq 0.2$.
The complex optical properties in the iron-arsenic planes have been examined
in the normal and superconducting states.  In the normal state, the free
carrier response is dominated by the electron pocket and may be described
by a simple Drude-Lorentz model.  In the generalized-Drude response for
$T\simeq T_c$ the frequency-dependent scattering rate shows clear kinks
at about $\simeq 12$~meV and 55~meV, suggesting scattering from bosonic
excitations.
Below $T_c$ the formation of a condensate is observed; however, less than 50\%
of the free carriers condense, suggesting this material is not in the clean limit.
Supporting this view is the observation that this material is observed to fall
close to the universal scaling line for a BCS dirty-limit superconductor in the
weak-coupling limit.
In order to properly model the optical conductivity below $T_c$ it is necessary
to introduce two superconducting energy gaps at $\Delta_1(0) = 3.1\pm 0.2$~meV
and $\Delta_2(0) = 7.4\pm 0.4$~meV.  This corresponds to either the gapping of
the electron and hole pockets, respectively, or the gapping of the electron
pocket by an anisotropic {\em s}-wave gap; both results are consistent with
the $s^\pm$ model.  The time evolution of the photoinduced reflectivity
change in the superconducting state yields almost identical values for the gaps.
These results indicate that there are two energy scales for the superconductivity in BaFe$_{1.85}$Co$_{0.15}$As$_2$ and that the pairing is likely not
phonon mediated.

%
%
\section{Experiment}
\begin{figure}[t]
\centerline{\includegraphics[width=7.5cm]{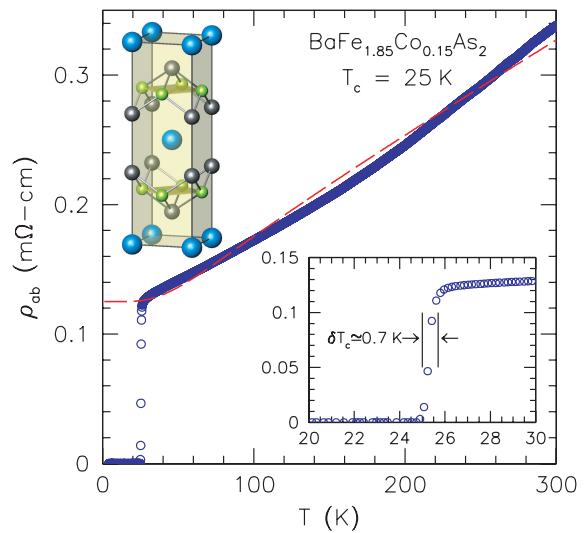}}
\caption{Temperature dependent {\em ab}-plane resistivity of a
BaFe$_{1.85}$Co$_{0.15}$As$_2$ single crystal with a superconducting
transition at $T_c =25$~K; the dashed curve is a Bloch-Gr\"{u}neisen
fit to the resistivity data (Sec.~IIIC).
Insets: The unit cell of BaFe$_2$As$_2$ in the tetragonal $I4/mmm$ setting;
detail of the resistivity in the region of $T_c$ illustrating the sharp
transition width $\delta T_c \simeq 0.7$~K.
%
}
\label{fig:trans}
\end{figure}

Large single crystals of the iron-arsenic superconductor
BaFe$_{2-x}$Co$_{x}$As$_2$ were grown by a self-flux method.\cite{li09}
Energy dispersive x-ray (EDX) microanalysis was used to determine that the
Co concentration $x=0.15$ was at the optimal level.  The resistivity in the
{\em ab}-planes was measured using a standard four-probe method.  The
temperature dependence of the resistivity of a single crystal of
BaFe$_{1.85}$Co$_{0.15}$As$_2$ is shown in Fig.~\ref{fig:trans}; the resistivity
displays a slight curvature before abruptly going to zero at $T_c \simeq 25$~K,
with a transition width of less than 0.7~K (inset of Fig.~\ref{fig:trans}).
%
%
The temperature-dependent reflectance was measured at a near-normal angle of
incidence from $\simeq 20$ to over 25,000~cm$^{-1}$ ($\simeq 2$~meV to 3~eV) on
a cleaved surface for light polarized in the {\em ab}-planes using {\em in situ}
evaporation technique.\cite{homes93}  This large frequency interval is needed to carry
out a reliable Kramers-Kronig analysis where extrapolations are supplied in
the limits for $\omega \rightarrow 0, \infty$.
For the transient reflectivity measurements, a Ti:sapphire laser system which can deliver
100-fs short pulses at an 80-MHz repetition rate, tunable at 800 nm (1.55~eV),  was used
as the source of both pump and probe pulses. A standard pump-probe setup is employed with
the pump beam having a spot-diameter of 60~$\mu$m and the time-delayed probe beam with
spot-diameter of 30~$\mu$m. The pump beam was modulated at 2~kHz with an optical chopper
and a lock-in amplifier was used to measure the transient reflectivity change, $\Delta{R}/R$,
of the probe beam.

%
%
\begin{figure}[t]
\centerline{\includegraphics[width=7.5cm]{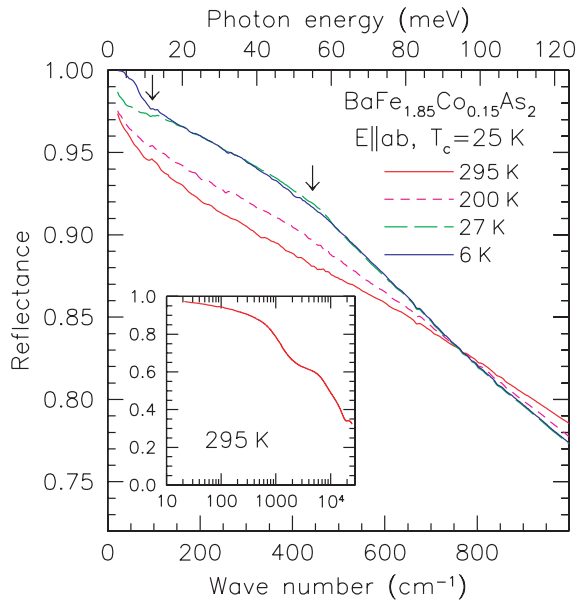}}
\caption{The temperature dependence of the reflectance of single-crystal
BaFe$_{1.85}$Co$_{0.15}$As$_2$ for light polarized in the {\em a-b} planes at
several temperatures above and below $T_c$.  There is a dramatic change in
the reflectance below $T_c$.   The "kinks" in low-temperature
normal-state reflectivity (27 K) at $\simeq 12$ and 55 meV (marked by the
two arrows at the inflection points in reflectivity) result from
strong scattering of the carriers with some underlying bosonic excitations
at these energy scales.
Insert: The reflectance over a wide frequency range at 295 K. }
\label{fig:reflec}
\end{figure}

%
%
\section{Results and discussion}
\subsection{Optical properties}
\subsubsection{Normal state}
The temperature dependence of the reflectance of BaFe$_{1.85}$Co$_{0.15}$As$_2$
for light polarized in the {\em ab}-planes is shown in the infrared region in
Fig.~\ref{fig:reflec} for several temperatures above and below $T_c$; the
reflectance at 295~K is shown over a much larger region in the inset.  In the
normal state, the reflectance at low frequency has a $R\propto 1 - \sqrt{\omega}$
response characteristic of a metal in the Hagen-Rubens regime.  In addition to this
generic response, just above $T_c$ at 27~K there are two inflection points in the
reflectance (indicated by arrows) at $\approx 12$ and 55~meV, which may result
from scattering of the carriers with underlying bosonic excitations.
Below $T_c$ the superconducting state displays a clear signature in the
reflectance.  However, the reflectance is a complex quantity consisting
of an amplitude and a phase, $\tilde{r} = \sqrt{R}\exp(i\theta)$.  Normally, only the
amplitude $R = \tilde{r} \tilde{r}^\ast$ is measured so it is not always intuitively
obvious what changes in the reflectance indicate.  For this reason, the complex
optical properties have been calculated from a Kramers-Kronig analysis of the
reflectance.\cite{dressel-book}

\begin{figure}[b]
\centerline{\includegraphics[width=7.5cm]{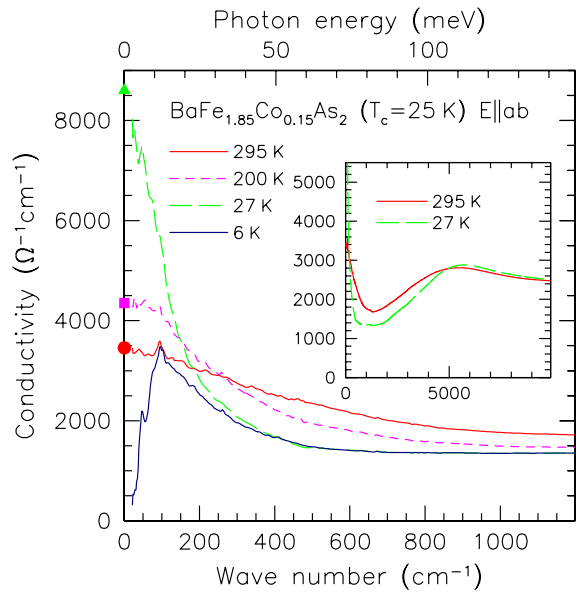}}
\caption{The real part of the optical conductivity of BaFe$_{1.85}$Co$_{0.15}$As$_2$
in the infrared region for light polarized in the {\em a-b} planes at several
temperatures above and below $T_c$.  The extrapolated values for $\sigma_{dc}
\equiv \sigma_1(\omega\rightarrow 0)$ agrees quite well with the dc transport data
from Fig.~\ref{fig:trans} at 295 ({\CIRCLE}), 200 ($\blacksquare$) and 27~K ($\blacktriangle$).  Below $T_c$ there is a dramatic loss of low-frequency
spectral weight due to the formation of a condensate.
Inset: The conductivity at 295 and 27~K over a much wider frequency range.}
\label{fig:sigma}
\end{figure}

%
%
The temperature dependence of the real part of the optical conductivity
$\sigma_1(\omega)$ of BaFe$_{1.85}$Co$_{0.15}$As$_2$ in the far-infrared
region is shown in Fig.~\ref{fig:sigma}.  At room temperature the conductivity
decreases slowly with increasing frequency giving way, not unlike the
high-$T_c$ cuprates,\cite{tu02} to a rather flat response in the
mid-infrared region followed by a pronounced peak at about 5200 cm$^{-1}$
(shown in the inset).  At room temperature a weak infrared-active $E_u$ mode
at about 94~cm$^{-1}$ is observed that involves in-plane displacements of the
Ba atoms,\cite{litvinchuk08} a second $E_u$ mode expected at 253~cm$^{-1}$ is
a weak feature observed only a low temperatures, suggesting that this mode may
be broadened due to disorder effects.\cite{akrap09}
As the temperature is reduced the low-frequency conductivity increases as a
result of the transfer of spectral weight from high to low frequencies, and
the low-frequency $E_u$ mode becomes more difficult to observe.
The spectral weight is defined simply as the area under the conductivity
curve over a given interval, $N(\omega, T) = \int_0^\omega
\sigma_1(\omega^\prime, T) d\omega^\prime$.
%
%
It is important to note that the values for the dc conductivity in
Fig.~\ref{fig:trans} at 295, 200 and 27~K are in good agreement with the
extrapolated values $\sigma_{dc} \equiv \sigma_1(\omega\rightarrow 0)$
in Fig.~\ref{fig:sigma}; the fact that $\sigma_{dc} \simeq \sigma_1(\omega
\rightarrow 0)$ establishes a connection with transport and illustrates
the self-consistent nature of this optical technique.

%
%
\begin{table}[tb]
\caption{The parameters for the Drude-Lorentz fits to in-plane optical
conductivity at 27~K, where $1/\tau_{D,j}$ and $\omega_{p,D;j}$ are the
scattering rate and plasma frequency for the $j$th Drude component (D$_j$),
and $\omega_k$, $\gamma_k$ and $\Omega_k$ are the frequency, width and
oscillator strength of the $k$th mode (L$_k$).  The top frame is the result
from considering two Drude and a single Lorentz oscillator.  The
second frame considers two Drude and two Lorentz components, while
the bottom frame considers a single Drude component and two Lorentz
oscillators.  All units are in cm$^{-1}$, unless otherwise indicated.}
\begin{ruledtabular}
\begin{tabular}{ccccc}
  Type & $\omega_{k}$ & $1/\tau_{D,j}$, $\gamma_k$ & $\omega_{p,D;j}$, $\Omega_k$ & $l$ (\AA ) \\
  \cline{1-5}
   D$_1$  &       &  113 &  6943 & 58 \\
   D$_2$  &       & 5720 & 19980 & 0.4\\
   L$_1$  &  5170 & 5730 & 22070 & \\
  \cline{1-5}
   D$_1$  &       &  126 &  7790 & 53\\
   D$_2$  &       &  608 &  1956 & 3.5\\
   L$_1$  &  1124 & 5560 & 18930 & \\
   L$_2$  &  5190 & 5780 & 22470 & \\
  \cline{1-5}
   D$_1$  &       &  126 &  7840 & 53 \\
   L$_1$  &  1016 & 5070 & 18250 & \\
   L$_2$  &  5200 & 5870 & 22980 & \\
\end{tabular}
\end{ruledtabular}
\label{tab:fits}
\end{table}

%
%
One technique to reproduce the optical conductivity assumes that because
these materials are multiband systems that the free-carrier response
can be described by separate contributions from the electron and hole
pockets, while the optical properties at high frequency are described by
bound excitations.  In a Drude-Lorentz model, this description results in
the linear combination of two Drude components.\cite{wu10a}  The
Drude-Lorentz model for the dielectric function $\tilde\epsilon(\omega) =
\epsilon_1(\omega) +i\epsilon_2(\omega)$ can be written as
$$
  \tilde\epsilon(\omega) = \epsilon_\infty -
  \sum_j{{\omega_{p,D;j}^2}\over{\omega^2+i\omega/\tau_{D,j}}}
   + \sum_k {{\Omega_k^2}\over{\omega_k^2 - \omega^2 - i\omega\gamma_k}},
$$
where $\epsilon_\infty$ is the real part of the dielectric function at high
frequency, $\omega_{p,D;j}^2 = 4\pi n_j e^2/m_j^\ast$ and $1/\tau_{D,j}$ are the
square of the plasma frequency and scattering rate for the delocalized (Drude)
carriers in the $j$th pocket, respectively; $\omega_k$, $\gamma_k$ and $\Omega_k$
are the position, width, and strength of the $k$th vibration or excitation.
The complex conductivity is $\tilde\sigma(\omega) = \sigma_1(\omega) +
i\sigma_2(\omega) = -i\omega [\tilde\epsilon(\omega) - \epsilon_\infty ]/4\pi$.

%
%
\begin{figure}[t]
\centerline{\includegraphics[width=7.5cm]{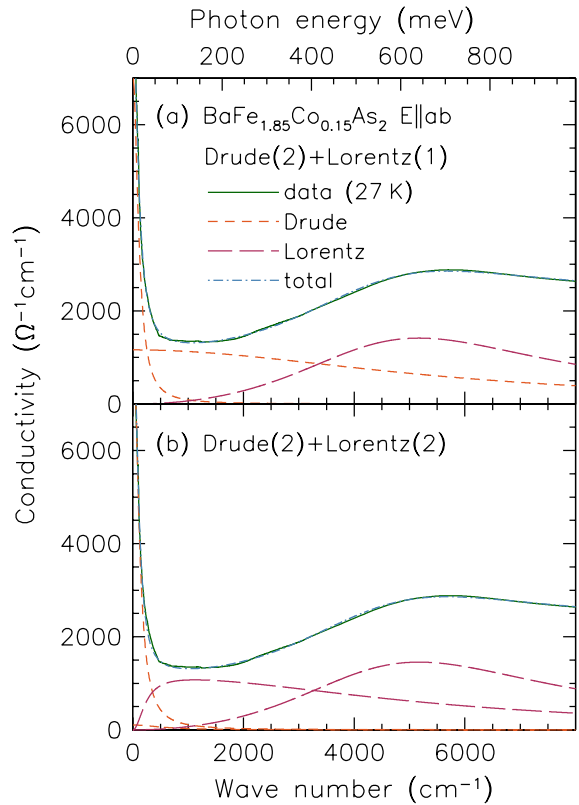}}
\caption{(a) The two-Drude and single Lorentz oscillator fit
to the data at 27~K over a wide frequency range.  The Drude
responses can be characterized by narrow and broad bands,
with a single Lorentz oscillator centered at $\simeq
5200$~cm$^{-1}$.
(b) The fit to the data at 27~K using a two Drude components
and two Lorentz oscillators.  A significant portion of the
spectral weight of the broad Drude component has been transferred
to the low-frequency Lorentz oscillator.  The results of the fits
are summamrized in Table~\ref{tab:fits}.
}
\label{fig:fits}
\end{figure}

%
%
The results of the two-Drude fit to the real part of the optical conductivity
just above $T_c$ at 27~K are shown in Fig.~\ref{fig:fits}(a) and summarized in
Table~\ref{tab:fits}.  The two Drude responses may be characterized as
``narrow'' ($1/\tau_{D} \simeq 113$~cm$^{-1}$) and ``broad'' ($1/\tau_{D}
\simeq 5700$~cm$^{-1}$); an additional Lorentz oscillator at $\simeq 5200$~cm$^{-1}$
has also been included as well as several others above 1~eV that are not
allowed to vary (not listed in Table~\ref{tab:fits}).
While the fit to the data is quite good, the scattering
rate for the broad Drude component is more than 50 times that of the narrow
component, which raises the issue as to whether this represents a mean-free
path $l_j = v_{F,j} \tau_j$ that is physically meaningful ($v_{F,j}$ is the
Fermi velocity and $\tau_j$ is the scattering time of the $j$th pocket).
%
%
There are a range of values in the literature for the Fermi velocities for
the electron and hole pockets in the iron-arsenic materials.  Based on
photoemission study of BaFe$_{1.85}$Co$_{0.15}$As$_2$ (Ref.~\onlinecite{sekiba09}),
we estimate that the Fermi velocity of the hole pocket to be $v_F \simeq
270$~meV$\cdot$\AA , or $v_F \simeq 6.5 \times 10^5$~cm/s.  This result is
less than the value determined from a band structure estimate of $v_F$
for the hole pocket in a related material ($\simeq 8 \times 10^6$~cm/s).\cite{singh08a}
However, it has been noted that band-structure results often have to be
renormalized to agree with photoemission results.\cite{neupane10}
The velocity of electron pocket is more difficult to calculate.  As an
estimate we will use the renormalization of the hole pocket Fermi velocity
to the band structure estimate of $v_F$ on the electron
pocket\cite{singh08a} to obtain $v_F \simeq 2 \times 10^6$~cm/s.
%
%
We associate the narrower Drude component with the electron pocket with
$\tau = 2.9 \times 10^{-13}$~s which leads to a mean-free path of
$l_e \simeq 58$~\AA , or about 15 unit cells ($a \simeq 4$~\AA\ in the
I4/mmm setting).  Associating the broad Drude component with
the hole pocket with $\tau \simeq 5.8 \times 10^{-15}$~s leads to a
mean-free path of $l_h \simeq 0.4$~\AA , which is less than the shortest
interatomic spacing; even a substantial increase in $v_F$ will not
alter this condition.  This places the mean-free path below the Mott-Ioffe-Regel limit\cite{mott72,gurvitch81,hussey10} and indicates that the conductivity
in this band is no longer metallic but is instead incoherent, and is best
described by a bound excitation, or a series of bound excitations.

%
%
This suggests that it is appropriate to include a low-frequency bound
excitation which my arise from localization effects or from interband
transitions.\cite{heumen09,hancock10}  We have therefore considered the
case of two Drude contributions as well as two low-frequency Lorentz
oscillators; as in the previous case oscillators above 1~eV are not fitted
quantities nor are their values changed from the previous instance.
The results of the fit to the data at 27~K are shown in Fig.~\ref{fig:fits}(b)
and are summarized in the middle frame of Table~\ref{tab:fits}.
The narrow Drude component at 27~K is only slightly broader than the same
component in the previous approach, resulting in a slightly shorter
mean-free path for the electron pocket $l_e \simeq 53$~\AA .  Much of
the spectral weight formerly associated with the broad Drude component
has shifted to shifted to a Lorentzian centered at $\simeq 1100$~cm$^{-1}$,
resulting in a dramatic reduction of the scattering rate, $1/\tau \simeq
600$~cm$^{-1}$ and a commensurate increase of the mean-free path
$l_h \simeq 3.5$~\AA\ or $l_h \simeq a$; this value is still close to the
Mott-Ioffe-Regel limit and it is debatable as to whether or not this
constitutes metallic transport.  While the uncertainties associated with
most of the fitted  parameters (determined from the covariance) are
$\lesssim 5$\% , the broad and weak Drude component is difficult to
fit and the errors associated with it are considerably larger,
$\approx 20${\%}.  If there are indeed two Drude components then
the electron pocket dominates and the hole pocket represents at best
a weak, possibly incoherent, background contribution as shown in
Fig.~\ref{fig:fits}(b).

%
%
For completeness we have also considered the case of a single Drude
component and two low-frequency Lorentz oscillators; the results of
this fit are summarized at the bottom of Table~\ref{tab:fits}.
With the exception of the absence of the second Drude component, the
results are essentially the same as Fig.~\ref{fig:fits}(b).  The
width of the narrow Drude component is unchanged and the slight
increase in strength reflects the fact that it has absorbed the
spectral weight formerly associated with the broad Drude component.
We conclude that despite the presence of multiple bands in this
material, for optimal doping the normal-state transport is
dominated by the single electron pocket.  The fitted values for
the Drude parameters at 295, 200 and 27~K are $\omega_{p,D} =
7840$~cm$^{-1}$ (or 0.972~eV) and $1/\tau_D = 304$, 233 and
126~cm$^{-1}$, respectively.

%
%
%
\begin{figure}[tb]
\centerline{\includegraphics[width=7.5cm]{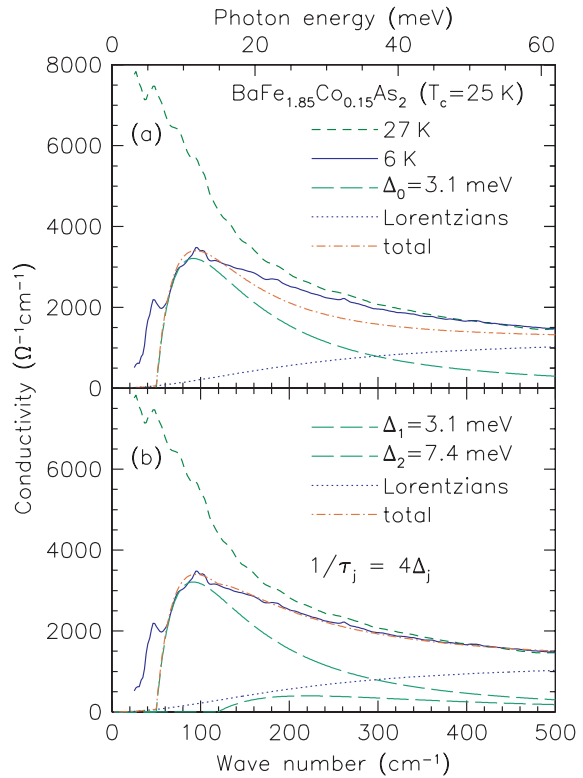}}
\caption{The in-plane optical conductivity of the iron-arsenic
superconductor BaFe$_{1.85}$Co$_{0.15}$As$_2$ ($T_c = 25$~K)
shown at 27 and 6~K (dashed and solid lines, respectively).
(a)  The optical conductivity with a single isotropic {\em s}-wave
gap of $\Delta_0 = 3.1$~meV with a scattering rate
$1/\tau=4\Delta_0$ is calculated for $T \ll T_c$ (long-dashed line)
and superimposed on the contribution from the bound excitations in
the mid-infrared (dotted line); the linear combination of
the two curves (dot-dash line) does not reproduce the data well
above $\approx 140$~cm$^{-1}$.
(b) The optical conductivity with isotropic {\em s}-wave gaps of
$\Delta_1 = 3.1$~meV and $\Delta_2 = 7.4$~meV  with the
scattering rates $1/\tau_j = 4\Delta_j$ is calculated for $T \ll T_c$
and superimposed on the Lorentzian contribution; the linear
combination of the three curves is in much better agreement with
the measured data.
}\label{fig:gaps}
\end{figure}

%
%
\subsubsection{Superconducting state}
Below $T_c$ there is a dramatic decrease in the low-frequency optical conductivity
due to the formation of a superconducting energy gap.  This ``missing area'' is
referred to as the spectral weight of the condensate $N_c$ and may be calculated
from the Ferrell-Glover-Tinkham sum rule\cite{ferrell58,tinkham59}
$$
  N_c \equiv N(\omega_c, T\simeq T_c) - N(\omega_c, T\ll T_c) = \omega_{p,S}^2/8.
$$
Here $\omega_{p,S}^2 = 4\pi n_s e^2/m^*$ is the square of the superconducting
plasma frequency and superfluid density is $\rho_{s0} \equiv \omega_{p,S}^2$;
the cut-off frequency $\omega_c\simeq 400$~cm$^{-1}$ is chosen so
that the integral converges smoothly.  The superconducting plasma frequency
has also been determined from $\epsilon_1(\omega)$ in the low frequency limit
where $\epsilon_1(\omega) = \epsilon_\infty - \omega_{p,S}^2/\omega^2$.  Yet another
method of extracting $\omega_{p,S}$ from $\epsilon_1(\omega)$ is to determine
$[-\omega^2\epsilon_1(\omega)]^{1/2}$ in the $\omega\rightarrow 0$ limit.\cite{jiang96}
All three techniques yield $\omega_{p,S} \simeq 5200\pm 400$~cm$^{-1}$, indicating that less
than one-half of the free-carriers in the normal state have condensed
($\omega_{p,S}^2/\omega_{p,D}^2 \lesssim 0.5$).
In a clean-limit system the scattering rate associated with the Drude
free-carrier response is much smaller than the size of the fully-formed
isotropic optical superconducting energy gap $1/\tau \ll 2\Delta_0$.
As a result the full spectral weight of the Drude component lies below
the optical gap $2\Delta_0$; for $T \ll T_c$ all of the free carriers
collapse into the condensate and $\omega_{p,S} \equiv \omega_{p,D}$.
In materials where the  size of the free-carrier scattering rate and the
optical gap are comparable, $1/\tau \gtrsim 2\Delta_0$ (the so-called dirty
limit) a significant amount of the Drude spectral weight lies above
$2\Delta_0$.  As a result for $T \ll T_c$ the free carriers are gapped
and there is no residual Drude component at low temperature; however, only
a fraction of the Drude spectral weight below the optical gap $2\Delta_0$
collapses into the condensate, leading to the condition that
$\omega_{p,S} < \omega_{p,D}$.  This is precisely what we observe,
implying that this material is not in the clean limit.
The superfluid density can also be expressed as an effective penetration
depth $\lambda_0 \simeq 3000\pm 300$~\AA , which is in good agreement with
other optical measurements\cite{heumen09,gorshunov10,kim10,nakajima10,wu10a,marsik10}
as well as estimates based on other techniques for materials with
similar cobalt concentrations.\cite{williams09,gordon09,luan10}
%

%
%
The low-frequency optical conductivity is shown in more detail in
Fig.~\ref{fig:gaps}.  As previously noted, below $T_c$ there is a dramatic
reduction of the conductivity below about 80~cm$^{-1}$ ($\approx 10$~meV),
which sets the energy scale for the superconducting energy gap $\Delta_0$
and the optical gap $2\Delta_0$ [here we adopt the convention that $\Delta_j
\equiv \Delta_j(0)$].  We note that based on this initial estimate it appears
that $1/\tau_D \gtrsim 2\Delta_0$, suggesting that the conductivity in this
material can be modeled using a Mattis-Bardeen approach\cite{mattis58}
for the contribution from the gapped excitations.\cite{zimmermann91}
This dirty-limit approach is consistent with
the observation that less than 50\% of the free carriers in the normal
state collapse into the condensate for $T \ll T_c$.  Note that in this
instance, ``dirty'' refers to scattering from disorder and/or electronic
correlations in addition to any possible impurity effects.  In
Fig.~\ref{fig:gaps}(a) the optical conductivity in the superconducting
state is described by a single isotropic gap $\Delta_0 \simeq 3.1$~meV
($T \ll T_c$) with a moderate amount of elastic scattering
[$1/\tau = 4\Delta_0$, which is approximately the previously determined
value of $1/\tau_D$ just above $T_c$] in linear combination with the
low-frequency tails of the mid-infrared Lorentzian oscillators.
While the leading-edge of the conductivity is described fairly
well, there is a noticeable disagreement above about 140~cm$^{-1}$.

%
%
%
To properly model the conductivity over the entire far-infrared region,
two isotropic gap features are considered at $\Delta_1 = 3.1$~meV and
$\Delta_2 = 7.4$~meV with $1/\tau_j = 4\Delta_j$ for
$T \ll T_c$.  In combination with the Lorentzian tails, the two-gap
scenario shown in Fig.~\ref{fig:gaps}(b) fits the data quite well.
The gap amplitudes determined here are in excellent agreement with
recent optical results on BaFe$_{2-x}$Co$_x$As$_2$ for a similar cobalt
concentration,\cite{fischer10,heumen09} as well as photoemission\cite{terashima09}
and tunneling results.\cite{massee09}  While the lower gap ratio
$2\Delta_1/k_{\rm B}T_c \simeq 2.9$ is close to the BCS weak coupling limit
of 3.5, the upper gap ratio $2\Delta_2/k_{\rm B}T_c \simeq 7.3$
is considerably larger.
While the data below about 40~cm$^{-1}$ is not reproduced exactly, there is a
considerable uncertainty associated with the conductivity in this spectral region.
This suggestion that the Fermi surface is completely gapped in the optimally-doped
material is in agreement with thermodynamic studies.\cite{tanatar10,reid10}

%
%
The examination of the normal-state concluded that while the
low-frequency optical properties were dominated by the electron
pocket, a weak contribution from the hole pocket might also be
present.  In in the formalism employed here, the inclusion of
a second gap necessarily implies a second band.
The calculated Drude response just above $T_c$
in Fig.~\ref{fig:gaps}(b) for the two bands (not shown) is
$\sigma_{1,j}(\omega) = \sigma_{0,j}/(1+\omega^2\tau_j^2)$, where
$\sigma_{0,1} \approx 7700$~$\Omega^{-1}$cm$^{-1}$ and $\sigma_{0,2}
\approx 850$~$\Omega^{-1}$cm$^{-1}$ for the bands associated with the
small and large gaps, respectively.  Given that $\sigma_{0,j} =
\omega_{p,D;j}^2\tau_j/60$, where $\omega_{p,D;j}$ is the plasma
frequency associated with the $j$th band, and $\omega_{p,D}
\simeq ( \omega_{p,D;1}^2+\omega_{p,D;2}^2 )^{1/2}$ and
recalling that $\tau_j = 4\Delta_j$, we estimate that $\omega_{p,D;1}
\simeq 6800$~cm$^{-1}$ and $\omega_{p,D;2} \simeq 3500$~cm$^{-1}$.
Once again it is the case that a single band dominates and
that over 75\% of the superconducting condensate originates from
the gapping of this band.

%
%
In the simple approach employed here, the two energy scales for the gaps
result from the gapping of two separate bands.  This view is consistent
with ARPES measurements\cite{ding08,khasanov09,evtushinsky09,koitzsch09,%
zabolotnyy09,terashima09} that indicate that both the electron
and hole pockets are likely completely gapped below $T_c$.  However,
it is clear that the coherent transport in the normal state arises
mainly from the electron pocket.  It is therefore also possible that
the two energy scales for the superconducting energy gaps observed
here are due to single anisotropic {\em s}-wave gap on the electron
pocket, which is favored within the $s^\pm$ model.  Because
there is some uncertainty in the optical measurements with respect
to the amplitude of the larger gap, time-domain measurements were
performed.

%
%
\begin{figure}[t]
\centerline{\includegraphics[width=8.6cm]{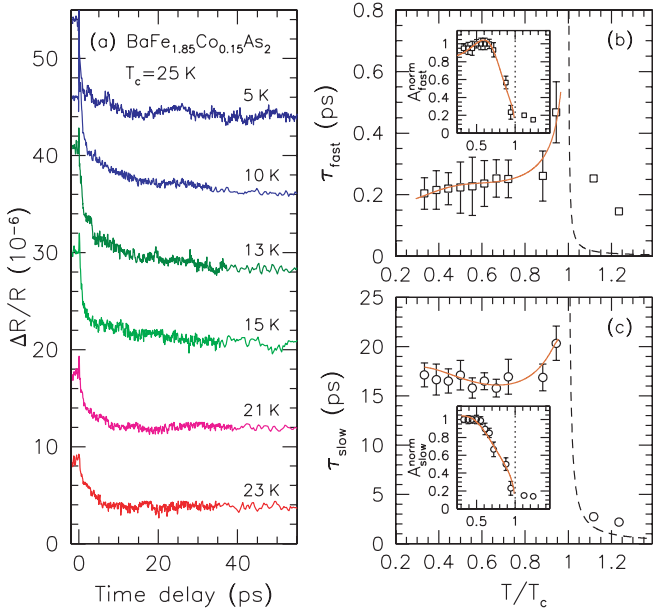}}
\caption{(a) The time evolution of the photo induced reflectivity change,
$\Delta{R}/R$ in BaFe$_{1.85}$Co$_{0.15}$As$_2$ as a function of temperature
(below $T_c$) at 800 nm (the spectra are offset for clarity).
(b) Temperature dependence of the relaxation time, $\tau_{fast}$. The inset
shows the normalized amplitude $A_{fast}$. Solid lines are fits to the RT
model, yielding $\Delta_2 \approx 7.4\pm 1.2$~meV;
(c) Temperature dependence of the relaxation time, $\tau_{slow}$.  The inset
shows the normalized amplitude $A_{slow}$. Solid lines are fits to the RT
model, yielding $\Delta_1 \approx 3.5\pm 0.4$~meV.}
\label{fig:time}
\end{figure}

%
%
%
\subsection{Time-domain spectroscopy}
The details of the superconducting gap properties have been investigated
in the time-domain.   Fig.~\ref{fig:time}(a) shows the time evolution of the
photo-induced reflectivity change, $\Delta{R}/R$ as a function of temperature
at 800~nm.  The data is fitted by $\Delta{R}/R = A_0 + A_{fast}\exp(-t/\tau_{fast})+
A_{slow}\exp(-t/\tau_{slow})$; $\Delta{R}/R$ shows a bi-exponential decay with relaxation
times $\tau_{fast} \simeq 0.2$~ps and $\tau_{slow} \simeq 15$~ps. These two decay
signals are the result of Cooper-pair recombination following the photo-excitation.
Figs.~\ref{fig:time}(b) and \ref{fig:time}(c) show the temperature dependence of the
relaxation times ($\tau_{fast}$ and $\tau_{slow}$) and the peak amplitudes are shown
in the insets ($A_{fast}$ and $A_{slow}$).  The fact that $\Delta{R}/R$ does not
return to its value at $t<0$ above about 45~ps is due to a negative component with a
long relaxation time ($\sim\,$ns), not a rise time. This negative component is a
common feature in photo-induced reflectivity measurements of complex metallic
systems involving magnetism such as colossal magnetoresistance materials\cite{ren08b}
and the high-$T_c$ cuprates;\cite{chia07} this component is not related to
the superconductivity.
The phenomenological Rothwarf-Taylor (RT) model\cite{rothwarf67} is used
to describe the quasiparticle dynamics
$$
  {{dn}\over{dt}} = I_0 + \beta N - Rn^2
$$
and
$$
  {{dN}\over{dt}} = -{{\beta N}\over 2} + {{Rn^2}\over 2} -(N-N_T)\gamma .
$$
Here $n$ and $N$ are the total density of quasiparticles and high-frequency bosons,
respectively, $\beta$ is the probability for pair breaking by high-frequency boson
absorption, $R$ is the bare quasiparticle recombination rate, $N_T$ is the density of
high-frequency bosons in equilibrium, and $\gamma$ is their decay rate.  We obtain
the normalized amplitude of the signal (to its low-temperature value), $A(T)$ via the
relation: $A(T) \propto (n_T -1)^{-1}$ , where the density of the thermally excited
quasiparticles, $n_T \propto \sqrt{\Delta(T)\,T} \exp[-\Delta(T)/T]$, with $\Delta(0)$
as a fitting parameter and $\Delta(T)$ obeying a BCS temperature dependence.  Moreover,
for a constant pump intensity and assuming $\gamma$ is temperature independent, we
obtain the temperature dependence of the relaxation time $\tau$ by the temperature
dependence of $n_T$:
$$
  \tau(T)^{-1} = \Gamma \left[ A(T) +b\sqrt{\Delta(T)\,T}\, \exp{\left( -{{\Delta(T)}/T}
                 \right)} \right] ,
$$
where $\Gamma$ and $B$ are fitting parameters. Since the temperature dependence of $A(T)$
and $\tau(T)$ are measured directly, we can accurately determine the values of the
superconducting gaps, $\Delta_1 = 3.5\pm 0.4$~meV and $\Delta_2 = 7.4\pm 1.2$~meV.
These gap values are in excellent agreement with the optical estimates determined in
the previous section.

%
%
\subsection{Electron-phonon coupling}
The temperature dependence of the {\em ab}-plane resistivity (Fig.~\ref{fig:trans}) has been
fit using the generalized Bloch-Gr\"{u}neisen formula\cite{gruneisen33,deutsch87,kong01,mamedov07}
$$
  \rho(T) = \rho_0 +\lambda_{ph} (m-1)
            \left( {{k_{\rm B}\Theta_D} \over {\omega_{p,D}^2}} \right)
            \left( {{T}\over{\Theta_D}} \right)^m
            J_m\left( {{T}\over{\Theta_D}} \right)
$$
where
$$
  J_m\left( {{T}\over{\Theta_D}} \right) = \int_0^{\theta_D/T} {{x^m e^{-x}}\over{(1-e^{-x})^2}} dx.
$$
In this instance the integer value $m=5$ implies that the resistance is due to the
scattering of electrons by phonons.  The Debye temperature $\Theta_D$ is set to
250~K (Ref.~\onlinecite{ni08}), and the plasma frequency has been previously
determined to be $\omega_{p,D} = 0.972\pm 0.05$~eV.  The fit, shown by the dashed
curve in Fig.~\ref{fig:trans} yields $\rho_0 \simeq 125$~m$\Omega$-cm and an
upper bound for the transport electron-phonon coupling constant $\lambda_{ph}
\simeq 0.2\pm 0.02$.  This experimental estimate of $\lambda_{ph}$ value is in
excellent agreement with the theoretical value of $\lambda \simeq 0.21$
(Ref.~\onlinecite{boeri08}) which yields a maximum $T_c \simeq 0.8$~K.
However, to get a more accurate estimate of the electron-boson coupling
constant from the resistivity, the temperature dependence of the underlying
bosonic spectra needs to be considered.

%
%
\subsection{Electron-boson coupling}
\begin{figure}[t]
\centerline{\includegraphics[width=7.5cm]{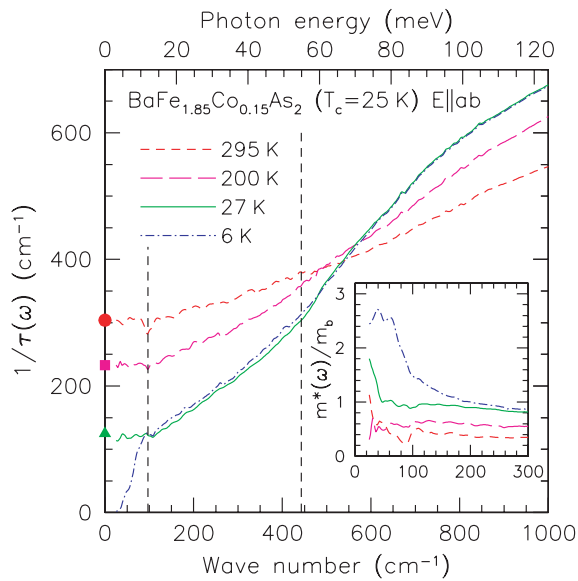}}
\caption{The in-plane frequency-dependent scattering rate of
 BaFe$_{1.85}$Co$_{0.15}$As$_2$ for several temperatures above and below $T_c$
 in the far-infrared region.  The values for $1/\tau_D$ are shown at 295 ({\CIRCLE}),
 200 ($\blacksquare$) and 27~K ($\blacktriangle$), respectively, where the
 scattering rate displays little temperature dependence.  The dashed lines
 at $\simeq 12$ and 55~meV indicate changes in the slope of $1/\tau(\omega)$.
 Inset: The frequency dependence of the effective mass.}\label{fig:tau}
\end{figure}

The kinks observed in the normal-state reflectance in Fig.~\ref{fig:reflec} at
$\simeq 12$ and 55~meV suggest scattering from underlying bosonic excitations.
In order to investigate this further the generalized Drude model has been considered
in which the scattering rate and the effective mass are allowed to adopt a frequency
dependence\cite{allen77,puchkov96}
$$
  {{1}\over{\tau(\omega)}} = {{\omega_p^2}\over{4\pi}} \,
  {\rm Re} \left[ {{1}\over{\tilde\sigma(\omega)}} \right]
$$
and
$$
  {{m^\ast(\omega)}\over{m_b}} = {{\omega_p^2}\over{4\pi\omega}} \,
  {\rm Im} \left[ {{1}\over{\tilde\sigma(\omega)}} \right] ,
$$
where the $m_b$ is the bandmass; $m^\ast(\omega)/m_b = 1+\lambda(\omega)$
and $\lambda(\omega)$ is a frequency-dependent electron-boson
coupling constant.  In this instance we set
$\omega_p \equiv \omega_{p,D}$ and $\epsilon_\infty = 4$
(although the choice of $\epsilon_\infty$ has little effect on the
scattering rate or the effective mass in the far-infrared region).
The temperature dependence of $1/\tau(\omega)$ is shown in
Fig.~\ref{fig:tau}, and the inset shows the temperature dependence of
$m^\ast(\omega)/m_b$.  In the normal state, in accord with the Drude
model, the scattering rate is constant below $\simeq 100$~cm$^{-1}$
and $1/\tau(\omega\rightarrow 0) \simeq 1/\tau_D$.  However, above
$\simeq 12$~meV there is an abrupt increase in the scattering rate
for [the large change in $1/\tau(\omega)$ for $T\ll T_c$ is due to the
formation of one or more superconducting energy gaps]; another kink is
observed at $\simeq 55$~meV.  Interestingly, the effective mass
enhancement is rather small $m^\ast(\omega\rightarrow 0)/m_b \simeq 2$,
indicating that the electron-boson coupling $\lambda\equiv
\lambda(\omega\rightarrow 0) \simeq 1$; this is significantly
smaller than the values of $\lambda \simeq 3 - 4$ that have recently been
reported.\cite{yang09,wu10b}

%
%

The electron-boson spectral function can
in be calculated from $1/\tau(\omega)$ using a maximum-entropy technique
based on Eliashberg theory.\cite{schachinger03}  Such an analysis has
recently been performed on a material with an almost identical cobalt
concentration and will not be repeated here; the peaks in the electron-boson
spectral function in that work\cite{wu10b} are observed at $\simeq 10$
and 45~meV, which is in good agreement with the kinks observed in the
present study in $1/\tau(\omega)$ at $\simeq 12$ and 55~meV.  The low-energy
peak is associated with the resonance peak in the spin excitation spectrum
obtained from inelastic neutron scattering data,\cite{inosov10} which
when considered with the relatively weak electron-phonon interaction in
this material, suggests that the superconductivity in this material may
be mediated by magnetic interactions.\cite{wu10b}  The high-energy kink
lies above the highest measured phonon energy;\cite{yildirim09} however,
muon-spin relaxation measurements\cite{carlo09} on a number
of iron pnictides suggest an antiferromagnetic energy scale of
$\approx 50$~meV indicating a possible magnetic origin for this feature.
It is not clear if this boson plays a role in the superconductivity
or not; however, we speculate that any mechanism that did couple to
the high-energy boson might display a significantly enhanced $T_c$,
suggesting that higher values for $T_c$ in this class of materials may
yet be possible.
%

%
%
\begin{figure}[b]
\centerline{\includegraphics[width=7.5cm]{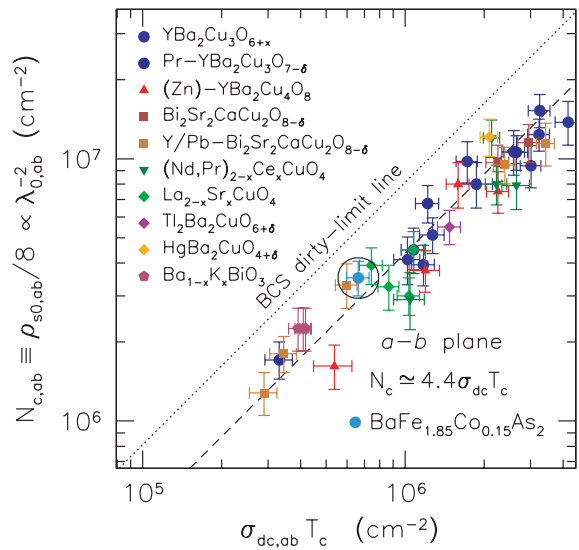}}
\caption{The log-log plot of the in-plane spectral weight of the superfluid
 density $N_c \equiv \rho_{s0}/8$ vs $\sigma_{dc}\,T_c$, for a variety of
 electron and hole-doped cuprates compared with the result for
 BaFe$_{1.85}$Co$_{0.15}$As$_2$.  The dashed line corresponds to the general
 result for the cuprates $\rho_{s0}/8 \simeq 4.4 \sigma_{dc}T_c$, while
 the dotted line is the result expected for a BCS dirty-limit superconductor
 in the weak-coupling limit, $\rho_{s0}/8 \simeq 8.1\, \sigma_{dc}T_c$.
}\label{fig:scaling}
\end{figure}

%
%
%
\subsection{Scaling of the superfluid density}
It has been demonstrated that a Uemura-type of scaling between the
superfluid density $\rho_{s0}$ and $T_c$ breaks down in the hole-doped
Ba$_{0.6}$K$_{0.4}$Fe$_2$As$_2$ material.\cite{ren08a}  However,
it has also been demonstrated that a number of the iron pnictide and
chalcogenide materials\cite{wu09b,homes10} the superfluid density
falls on a recently proposed empirical scaling relation for the
cuprate superconductors,\cite{homes04,homes05} shown by the
dashed line in Fig.~\ref{fig:scaling},
$$
 \rho_{s0}/8 \simeq 4.4\, \sigma_{dc}\, T_c .
$$
From the estimate of $\sigma_{dc} \equiv \sigma_1(\omega \rightarrow 0)
\simeq 8000\pm 400$~$\Omega^{-1}$cm$^{-1}$ for $T \gtrsim T_c$ (determined from
Fig.~\ref{fig:sigma}, as well as Drude-Lorentz fits), and the previously
determined value of $\rho_{s0} = 27\pm 3 \times 10^6$~cm$^{-2}$,
we can see that BaFe$_{1.85}$Co$_{0.15}$As$_2$ also falls on this scaling
line in the region of the moderately-underdoped cuprates.
The dotted line in Fig.~\ref{fig:scaling} is the calculated result for a
a BCS dirty-limit superconductor in the weak-coupling limit in which the
numerical constant in the scaling relation is calculated to be slightly
larger,\cite{homes05} $\rho_{s0}/8 \simeq 8.1 \sigma_{dc}\, T_c$.  This
scaling behavior is consistent with the observation that this material
is not in the clean limit.

%
%
\section{Conclusions}
The optical and transport properties of the iron-arsenic superconductor
BaFe$_{1.85}$Co$_{0.15}$As$_2$ ($T_c = 25$~K) have been measured over a
wide temperature and frequency range.  A Bloch-Gr\"{u}neisen analysis of
the resistivity indicates that the electron-phonon coupling $\lambda_{ph}\simeq 0.2$
is fairly weak, suggesting that the superconductivity in this material is not
mediated by lattice vibrations.  The optical properties in the normal state are
dominated by the electron pocket and may be described as a weakly-interacting
Fermi liquid, or Drude model, with $\omega_{p,D} \simeq 7840$~cm$^{-1}$ and
$1/\tau_D \simeq 126$~cm$^{-1}$ just above $T_c$ at 27~K.
The frequency-dependent scattering rate displays kinks at $\simeq 12$ and 55~meV
that correspond to peaks in the electron-boson spectral function, both of which
are believed to be magnetic in origin, and suggest that the pairing in this
material may be mediated by magnetic interactions.
In the superconducting state for $T \ll T_c$ the superconducting plasma frequency
$\omega_{p,S} \simeq 5200$~cm$^{-1}$, which corresponds to an effective penetration
depth of $\lambda \simeq 3000$~\AA .  This indicates that only about 50\% of the
free-carriers in the normal-state condense into the condensate, suggesting this
material is not in the clean limit.  In agreement with the results of other workers,
this material is observed to fall on the universal scaling line for a
BCS dirty-limit superconductor, and is also close to many underdoped cuprates.
The energy scales observed for the superconducting energy gaps
$\Delta_1 \simeq 3.1$~meV and $\Delta_2 \simeq 7.4$~meV correspond to
either the gaps on the electron and hole pockets, respectively, or the
minimum and maximum values for an anisotropic {\em s}-wave gap on the
electron pocket, in accord with the $s^\pm$ model. The spectrum of
excitations appears to be fully gapped, suggesting an absence of nodes
at optimal doping in this material.

%
%
\begin{acknowledgements}
We would like to thank A. Akrap, J. L. Birman,  G. L. Carr,
A. V. Chubukov, K. Felix, D. H. Lee, I. Mazin, P. Richard,
E. Schachinger, D. J. Singh and H. Yang and for helpful
discussions, and J. P. Carbotte for performing an inversion
of the optical data.
Work was supported by the National Science Foundation, and the National
Science Foundation of China.  Work at Brookhaven National Laboratory was
supported in part by the Office of Science, U.S. Department of
Energy under Contract No. DE-AC02-98CH10886.
\end{acknowledgements}
%

%
%
%

%

\end{document}